\documentclass{IOS-Book-Article}

\usepackage{mathptmx}
\usepackage{soul}\setuldepth{article}
\usepackage{graphicx}
\usepackage{url}
\graphicspath{{../pdf/}{../jpeg/}}
\DeclareGraphicsExtensions{.pdf,.jpeg,.png}
\usepackage{float}

\def\hb{\hbox to 11.5 cm{}}

\begin{document}

\pagestyle{headings}
\def\thepage{}
\begin{frontmatter}              

\title{The Telehealth Chain: a protocol for secure and transparent telemedicine transactions on the blockchain}

\markboth{}{June 2023\hb}

\author[A,B]{\fnms{Syed Sarosh} \snm{Mahdi}},
\author[C]{\fnms{Zaib} \snm{Ullah}}
\author[D]{\fnms{Gopi} \snm{Battineni}},
\author[B]{\fnms{Muneer Gohar} \snm{Babar}}
and
\author[B]{\fnms{Umer} \snm{Daood}}

\address[A]{Athena Center for Advanced Research in Healthcare, Camerino, Italy }
\address[B]{Division of Clinical Oral Health Sciences, School of Dentistry, International Medical University, Kuala Lumpur, Malaysia.}
\address[C]{Universita Telematica Giustino Fortunato, Benevento, Italy}
\address[D]{The clinical research center, School of Medicinal and Health Products Sciences, University of Camerino, Camerino, 62032, Italy.}

\begin{abstract}
Background:  Blockchain technology provides a secure and decentralized platform for storing and transferring sensitive medical data, which can be utilized to enable remote medical consultations. This paper proposes a theoretical framework for creating a blockchain-based digital entity to facilitate telemedicine services.
The proposed framework utilizes blockchain technology to provide a secure and reliable platform for medical practitioners to remotely interact with patient transactions. The blockchain will serve as a one-stop digital service to secure patient data, ensure privacy, and facilitate payments. The proposed framework leverages the existing Hyperledger Fabric platform to build a secure blockchain-assisted telemedicine platform.
\end{abstract}

\begin{keyword}
Blockchain Technology \sep Telemedicine \sep Hyperledger Fabric \sep Medical Records \sep Data Storage\sep Medical chain\sep IBM Health pass\sep Smart Contract\sep Cryptocurrency \sep 
\end{keyword}
\end{frontmatter}

\section{Introduction}\label{intro}
Blockchain Technology is offering innovative solutions to various complex problems e.g., smart healthcare, advanced transportation system, etc., in smart city development \cite{ullah2023blockchain}.
The current Covid-19 pandemic has demonstrated the significance of a secure, dependable, and robust healthcare industry response \cite{azim2020blockchain,nguyen2021blockchain,dar2022blockchain}. The pandemic has shed light on the importance of telehealth and telemedicine technology as it enables communication between physicians and patients without the need for face-to-face encounters, reducing the risk of infection transmission. Telemedicine services such as JD Health, Teledoc Health, and Rush University medical center have seen a surge in demand and rapid adoption during the current pandemic. These services have proven essential in combating Covid-19 spread \cite{chamola2020comprehensive}. 

However, the widespread adoption of telemedicine has also highlighted concerns about security, privacy, and transparency in telemedicine transactions. As patients and doctors interact remotely, there is a need for a secure and transparent platform to ensure that patient data is protected and transactions are executed efficiently. Sensitive medical records still lack a foul-proof security system which leaves them vulnerable to data breaches and hacks. 
Ownership of medical records is another major issue that arises with telemedicine. With the increased use of telemedicine, patient data is being collected and stored electronically. This data includes medical records, test results, and personal information. The question of who owns this data is a contentious issue. 

Traditionally, medical records were owned by healthcare providers, but patients were granted access to their records upon request. With the rise of telemedicine, patients are now collecting and storing their own data electronically. This has led to questions about who has control over this data. In the United States, the Health Insurance Portability and Accountability Act (HIPAA) outlines the rules regarding the ownership of medical records. Under HIPAA, healthcare providers own the medical records, but patients have the right to access and control their own medical information.Regenerate response. Blockchain technology can help to address the issue of ownership of medical records by providing a decentralized platform where patients can securely store and control their own medical data. By using blockchain technology, patients can control access to their medical data, ensuring that their information remains private and secure. In addition, patients can grant access to their medical data to healthcare providers as needed, ensuring that they receive the best possible care.

In 2008 a researcher named Satoshi Nakamoti developed the cryptocurrency called Bitcoin, which used blockchain technology and offered a secure, verifiable, and hack-resistant protocol for recording data \cite{nakamoto2008bitcoin}. Currently, more than 9000 cryptocurrencies are trading in the crypto market, offering various solutions on the blockchain. Most of these currencies are based on the Ethereum blockchain, another blockchain technology introduced in 2014 by Vittalik Butterin. The project we are proposing ( Telechain) will incorporate Filechain to deliver Telemedicine based interventions.

The article is organized as follows: 
Section \ref{relatedworks} delineates the related research work regarding blockchain and Hyperledger technology applications in healthcare. Section \ref{technicalbackground} provides detailed technical background and procedure of how to implement the proposed project.  Section \ref{smartcontract} discusses smart contracts and their implementation in a very detailed manner.  Section \ref{challenges} provides the advantages and possible challenges of using Hyperledge fabric-based blockchain technology for healthcare solutions. Finally, Sections \ref{conc} conclude the article.

\section{Related Works}\label{relatedworks}

Several Healthcare companies and organizations have taken up Hyperledger to secure their data as it has tremendous potential in the healthcare setting \cite{HyperledgerHealth}. Blockchain facilitates decentralizing data, without violating the security of sensitive healthcare data of patients.  The patient can seamlessly use their own credentials and signatures alongside the signatures of the hospital to log in to his/her own medical information to be used in the treatment. Through the blockchain model, the patient has complete control over their private medical data and only allows information that they deem essential to be shared with physicians \cite{ma2019privacy, gao2023data}.

The blockchain solution is also effective as it dispenses with the expensive process of retaining medical histories and records in the hospital. Blockchain technologies are also a solution against counterfeit drugs that pose a massive threat and burden to the pharmaceutical and medical industries \cite{haq2018blockchain}. Blockchain technologies can be used to keep counterfeit medicines out of the supply chain. Blockchain technology has great potential for supply chain management in the healthcare industry. Blockchain technologies make the recall of counterfeit drugs smooth and eventually reduce financial losses as well as improve safety and service delivery to customers \cite{pandey2021securing}. The following major healthcare players are leveraging the power of Hyperledger fabric to deliver health care.

 Change Healthcare is one of the largest healthcare organizations in the US and is utilizing the potential of Hyperledger fabric to design a safe \& private network to exchange healthcare data. The network will allow physicians and other healthcare workers to share patient data in a safe and secure manner, which in turn will reduce costs and enhance. the quality of care. The implementation of Hyperledger for change healthcare only took a few months in 2018 and they were able to process 50 million transactions per day, through the Hyperledger network \cite{ChangeHealthcare}. Similarly, IBM, Maersk, and other global players have adopted the Hyperledger fabric due to its secure and efficient network

IBM has built its IBM digital health pass on Hyperledger fabric. The service offers a multi-credential authenticator that companies can utilize to accomplish and implement their authentication and verification guidelines for Covid-19 vaccination status in a manner that satisfies both the privacy concerns of the individual and the public health requirements mandated by the organization or other relevant healthcare authorities \cite{IBMHealthpass}.  Using the IBM pass, multiple forms of Covid-19 health credentials can be verified seamlessly. The pass can verify the likes of EU Digital certificates, Smart Health cards the IBM Digital health pass, and the good health pass.

Another key industry player utilizing the power of Hyperledger fabric is the health consortium called Avaneer Health \cite{Hyperledger21, Hyperledger22}. Avaneer has created a healthcare network leveraging Hyperledger fabric blockchain solutions which are being exploited by various stakeholders of the consortium to increase collaboration and streamline the healthcare experience, hence leading to improved efficiency and administration \cite{Hyperledger23}. People using the Avaneer network can access a wide variety of blockchain technologies for securely handling and overseeing healthcare solutions and providing innovative healthcare products. Avaneer network includes 80 million covered individuals and 14 million patient visits annually. These are only a few examples of a wide variety of healthcare companies and organizations harnessing the power of the Hyperledger Fabric blockchain network in delivering healthcare solutions \cite{russo2022enabling}. KRYPC pharmaceutical delivery supply chain solution, Synaptic health alliance, and Maersk are among the many other key players using this secure and efficient solution \cite{Krypc23}.

\section{Technical Background}\label{technicalbackground}

Blockchain Technology:
Blockchain is a data decentralization technology based on a distributed digital ledger that records transactions in an expanding chain of unmodifiable blocks linked by cryptographic hashes. Typically, the user will start by looking for a transaction, which could involve smart contracts, data records, cryptocurrency transfer, or other types of data transfer \cite{Binance23, Coinmarketlinks}. The user then signs the transaction with his private key (similar to a digital signature), which allows other users to verify the transaction's authenticity using the public key. The public key is generated from the private key using a blockchain-verifiable computational method known as elliptic curve multiplication \cite{antonopoulos2014mastering}. The transaction is then broadcast to the entire peer-to-peer network of nodes, and miners choose a subset of the transactions to form a new block.
\subsection{How blockchain works in Telehealth systems }
Telehealth enables medical care experts to screen, diagnose, and treat patients from a distance by providing low-cost services, thereby limiting patient access and opening new doors in technology, reducing the workforce, and mitigating the risk of doctor exposure, staff, or patients during pandemics such as COVID-19 \cite{ismail2023cloud}. Furthermore, telehealth uses digital data and information and communication technologies (ICT) to help patients with specific illnesses through developed self-care and virtual education to support systems \cite{stowe2010telecare}. 

Before blockchain integration for telehealth, centralization was a significant barrier for telemedicine frameworks. In any case, centralized frameworks consistently represent the risk of a single point of failure and are highly susceptible to various outer and inner information breaks, threatening system reliability. Blockchain technology now effectively addresses such critical issues \cite{chamola2020comprehensive}. It employs a distributed engineering approach to dealing with the common record of medical records. All recorded duplicates are adjusted and checked with each blockchain-connected hub. Blockchain addresses key issues by following drugs and medications throughout the supply chain, tracking infected patient locations, verifying doctors' accreditations, and securing health records. 
\begin{figure}[H]
\centering
  \includegraphics[width=14cm, height=10cm]{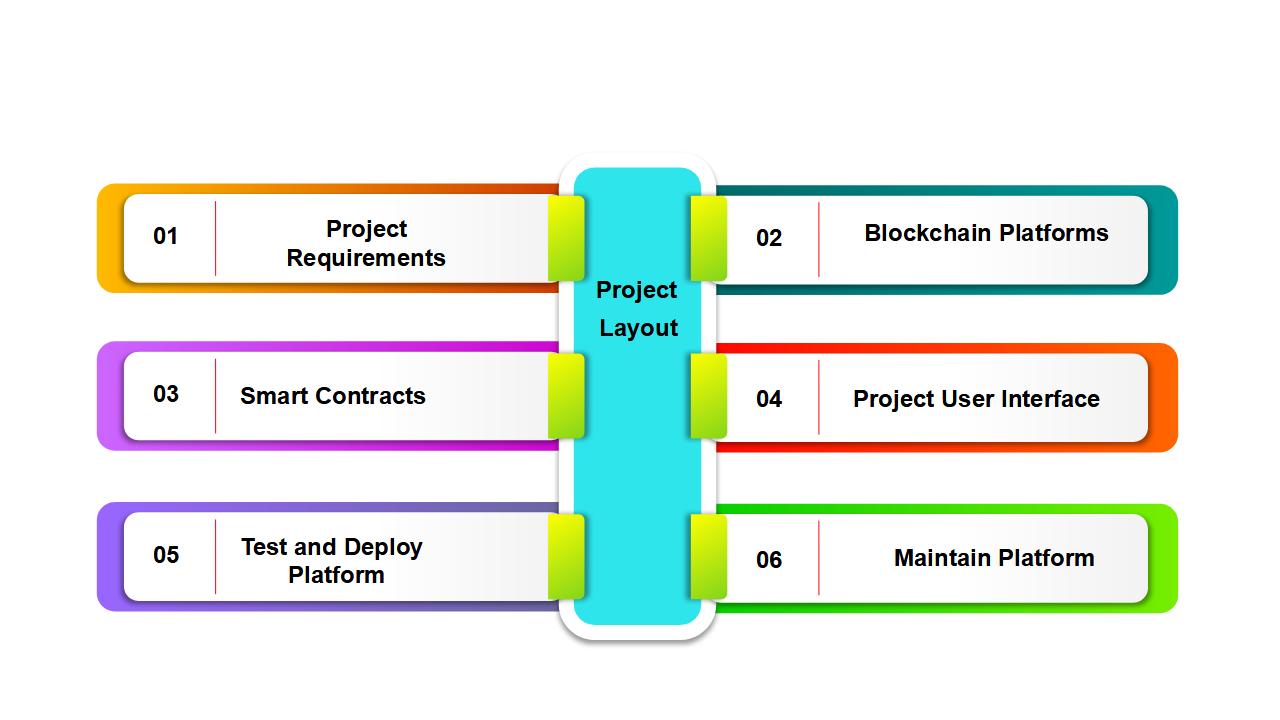}
  \caption{ Project requirements.}
  \label{fig:BCLayout1}
\end{figure}
\begin{figure}[H]
\centering
  \includegraphics[width=0.95\textwidth]{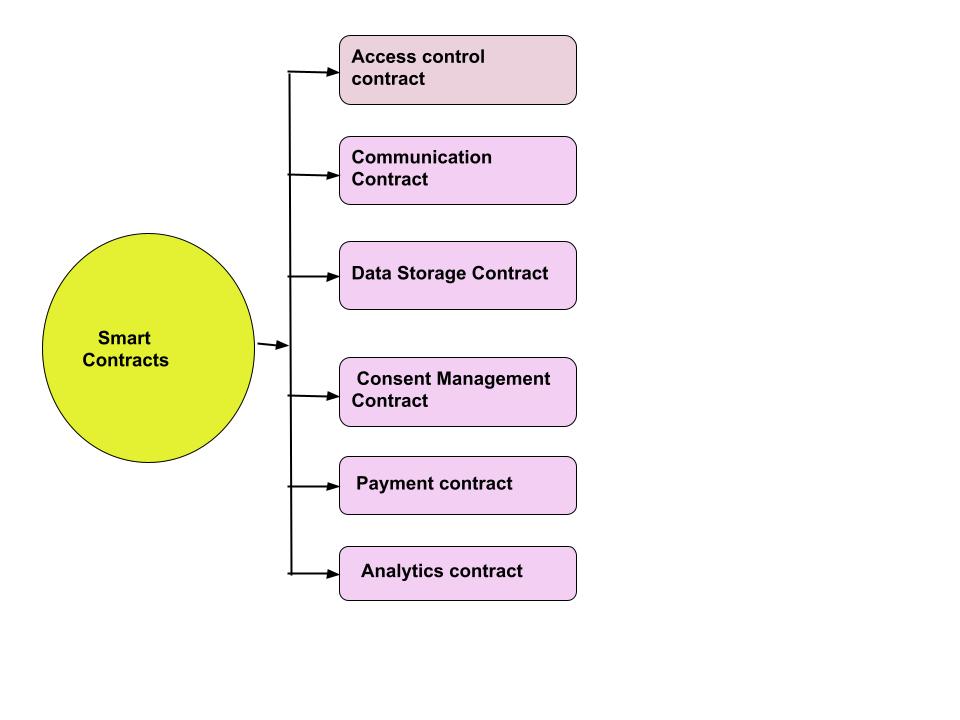}
  \caption{ Set of smart contracts required to develop the Tele-health chain.}
  \label{fig:BCLayout1}
\end{figure}

\subsection{Step-by-step procedure of blockchain functionality in Telehealth }
The proposed framework is comprised of many steps and expertise e.g., an understanding of the Hyperledger platform, its programming language, and a solid understanding of smart contracts development.
Implementing a framework that utilizes blockchain technology to provide a secure and reliable platform for medical practitioners to remotely interact with patients involves several steps. Here is an overview of the necessary steps:
\begin{itemize}

 \item Define the requirements: The first step is to define the requirements of the platform. This involves identifying the key features and functionalities needed to support remote interaction between medical practitioners and patients, such as secure communication channels, data encryption, and access control mechanisms.

 \item Choose the blockchain platform: The next step is to choose the appropriate blockchain platform to use. Considerations include the scalability of the platform, the consensus algorithm used, and the level of privacy and security provided. Hyperledger Fabric, Ethereum, and Corda are popular blockchain platforms used in healthcare.

 \item Develop smart contracts: Once the blockchain platform is chosen, the next step is to develop smart contracts that will govern the interactions between medical practitioners and patients. This involves defining the rules and procedures that will govern access to patient data, communication channels, and other key aspects of the platform.

 \item Develop the user interface: The user interface is the component that allows medical practitioners and patients to interact with the platform. This involves developing web or mobile applications that enable secure communication and access to patient data.

 \item Test and deploy the platform: Once the development is completed, it is essential to test the platform thoroughly to ensure that it meets the requirements and is secure. After testing, deploy the platform to the target environment, which may involve integration with other systems and compliance with regulatory requirements.

 \item  Maintain and update the platform: Finally, it is important to maintain and update the platform to ensure that it remains secure, reliable, and meets the changing needs of medical practitioners and patients. This may involve implementing new features, fixing bugs, and performing regular security audits.

\end{itemize}

\subsection{Steps for Deploying Telechain on Hyperledger Fabrics for Secure and Transparent Telemedicine Transactions}

To efficiently use the proposed project by patients and practitioners deployed using Hyperledger Fabric blockchain technology, the following are the necessary steps to carry out:
\begin{itemize}
    \item Understand the project necessities: Educate users on the detailed requirements and goals of the e-health project. Specify the functionalities and characteristics that the project contributes to patients and practitioners.
  \item Onboard patients and practitioners: Design a strategy for onboarding patients and practitioners onto the e-health platform. This may include signup, identity confirmation, and authorization techniques to assure the safety and privacy of user data.
  \item Familiarize users with blockchain technology: Supply academic resources to patients and practitioners to enable them to understand the basics of technology and how it improves the e-health system's security, clarity, and data integrity.
  \item Access the e-health platform: Patients and practitioners should have a user-friendly interface to access the e-health platform. Design a web or mobile application that permits users to log in, view their profiles, and access the available functionalities.
  \item User authentication and authorization: Implement a secure authentication approach to affirm the identity of patients and practitioners. Employ digital signatures, public-key infrastructure, or other cryptographic procedures to authenticate users and handle access control.
  \item Sustain patient records on the blockchain: Employ the Hyperledger Fabric blockchain to securely store and organize patient data. Describe the data structure for medical records, guaranteeing privacy and adherence with applicable decrees.
  \item Facilitate secure data sharing: Develop a permissioned network on Hyperledger Fabric that permits authorized practitioners to access and share patient records securely. Design fine-grained access management mechanisms to guarantee only relevant data is shared among allowed partakers.
  \item Enable transactions and payments: Design smart contracts on Hyperledger Fabric to automate payment processes. This can simplify billing, insurance claims, and other financial interactions between patients, practitioners, and other involved parties.
  \item Ensure privacy and data protection: Develop appropriate data privacy and protection procedures, such as data encryption, to protect sensitive patient information. Adhere to appropriate data protection regulations like GDPR (General Data Protection Regulation) or local privacy laws.
  \item Monitor and audit the blockchain network: Develop monitoring techniques to track the activity and efficiency of the Hyperledger Fabric network. Consistently audit the blockchain to guarantee data integrity and specify any irregularities or suspicious activities.
  \item Provide user support and feedback tools: Develop channels for users to pursue support or deliver feedback on the e-health platform. Show quick service to settle any issues or problems raised by patients and practitioners.
  \item Constantly enhance and update the system: Regularly assess the e-health platform's implementation, security, and user experience. Integrate user feedback, address specified deficiencies, and release updates or new attributes to improve the system's functionality and usability.
    
\end{itemize}

\section{Smart Contracts}\label{smartcontract}
Smart contracts and Hyperledger Fabric are interrelated, offering a powerful blend for secure and transparent transactions. With smart contracts, predefined rules and conditions ensure trust and efficiency in the Hyperledger Fabric network. This ingenious technology revolutionizes enterprises by lowering costs, eliminating intermediaries, and enabling a decentralized ecosystem of trust and confidence.

The set of smart contracts required to operate the above framework will depend on the specific requirements of the platform and the blockchain platform used \cite{khan2021blockchain}. However, here are some examples of smart contracts that could be used:
\begin{itemize}

\item Access control contract: This contract manages access control for medical practitioners and patients and enforces the rules for accessing patient data and communication channels.

\item  Communication contract: This contract manages secure communication channels between medical practitioners and patients and ensures that all communications are encrypted and authenticated.

\item Data storage contract: This contract manages the storage of patient data on the blockchain and ensures that all data is encrypted and immutable.

\item Consent management contract: This contract manages patient consent for accessing and sharing medical data and ensures that all patient data is used in accordance with applicable laws and regulations.

\item Payment contract: This contract manages payments for medical services and ensures that all payments are made securely and transparently.

\item  Analytics contract: This contract manages the analysis of patient data to generate insights and improve medical care while ensuring that patient privacy is protected.
\end{itemize}
These are just a few examples of the types of smart contracts that could be used in a blockchain-based platform for remote medical interactions. The specific set of contracts required will depend on the specific needs and requirements of the platform.

\subsection{Layout of  Smart Contracts Development}
In the following, we present the necessary steps inside the block diagrams to be accomplished for the development of the aforementioned smart contracts, to be specifically used on the Hyperledger platform \cite{merlec2021smartbuilder, yu2019comparison}.
 \begin{itemize}
     \item Access control contract: This contract manages access control for medical practitioners and patients and enforces the rules for accessing patient data and communication channels. The block diagram shown in Figure \ref{fig:AccessSC} portrays Access Control smart contract. The contract is composed of Grant Access, Approve Access, and Helper Method functions. Each respective function is further explained with detailed sub-operations to effectively fulfill assigned tasks and provide access to both patient and practitioner.

     \begin{figure}[H]
\centering
  \includegraphics[width=0.99\textwidth]{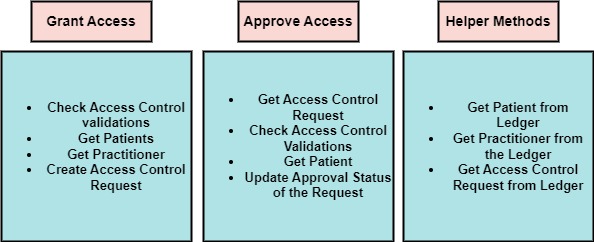}
  \caption{ Access Control Smart Contracts.}
  \label{fig:AccessSC}
\end{figure}

\item Communication contract:
The block diagram in figure \ref{fig:commSC} illustrates the Communication smart contract. The contract consists of Send Message, Receive Message, and Helper Method functions to ensure secure communication between patient and practitioner. Each function further performs assigned tasks e.g., data encryption, decryption, and writing it on blockchain ledgers in the form of data blocks.
     \begin{figure}[H]
\centering
  \includegraphics[width=0.99\textwidth]{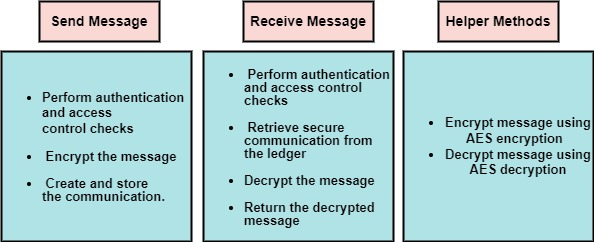}
  \caption{ Communication Smart Contracts.}
  \label{fig:commSC}
\end{figure}
\item Data storage contract:
The block diagram in figure \ref{fig:dscontract} depicts a data storage smart contract. The contract contains sub-functions like Store Data, Retrieve Data, and Helper Method to ensure that patients' and practitioners' data are securely stored on ledgers. Additionally, each sub-function conducts given tasks such as data encryption, decryption, storing and retrieving data, etc., as shown in the respective blocks.
 \begin{figure}[H]
\centering
  \includegraphics[width=0.99\textwidth]{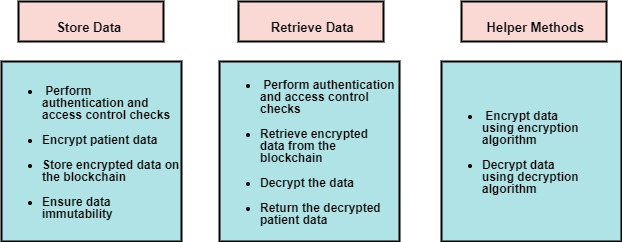}
  \caption{ Data storage smart contracts.}
  \label{fig:dscontract}
\end{figure}
\item Consent management contract:
The block diagram in figure \ref{fig:consentMSC} depict a consent management smart contract. The contract consists of sub-functions like Grant Consent, Revoke Consent, and Helper Method to ensure patient and practitioner data privacy. Each sub-function further accomplishes allocated tasks as shown in the respective blocks.
    \begin{figure}[H]
\centering
  \includegraphics[width=0.99\textwidth]{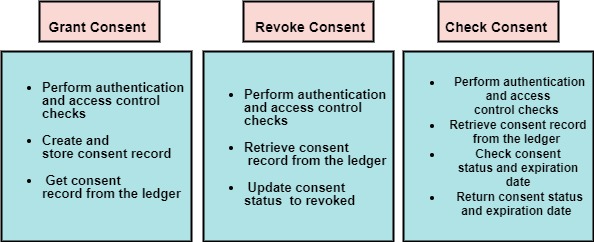}
  \caption{ Consent management smart contract.}
  \label{fig:consentMSC}
\end{figure}
\item Payment contract:
The block diagram in figure \ref{fig:paymentSC} depicts a payment smart contract. The payment contract includes sub-functions like Make Payment, Refund Payment, Check Payment Status, and Helper Method to efficiently and securely process payment-related  transactions. Each sub-function further conducts allocated assignments as shown in the respective blocks.
 \begin{figure}[H]
\centering
  \includegraphics[width=0.99\textwidth]{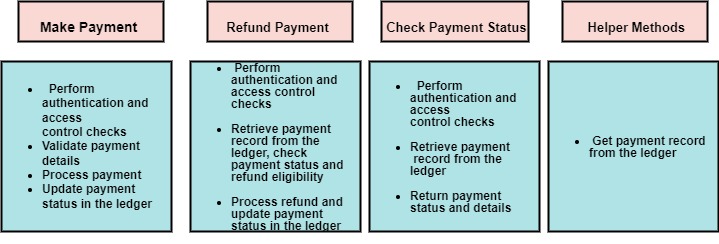}
  \caption{ Payment Smart Contracts.}
  \label{fig:paymentSC}
\end{figure}
\item Analytics contract:
The block diagram in figure \ref{fig:AnalticSC} depicts a data analytics smart contract. The contract includes sub-functions like Analyze Data, Get Insights and Helper Method. The contract investigates patient data, and develops insights, and patient data privacy protection is cared for during the process. The proposed block diagram can be further improved during its implementation phases. 
 \begin{figure}[H]
\centering
  \includegraphics[width=0.99\textwidth]{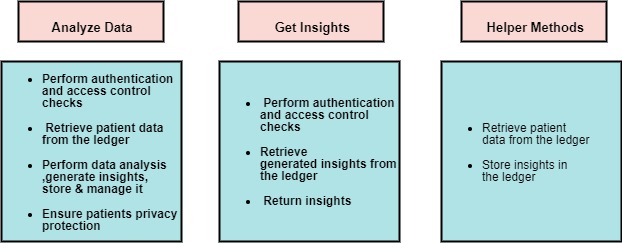}
  \caption{ Data Analytics Smart Contracts.}
  \label{fig:AnalticSC}
\end{figure}
 \end{itemize}

\section{ Challenges, Pros, and Cons} \label{challenges}
Using Hyperledger Fabric for healthcare undertakings presents both challenges and advantages. On the positive side, Hyperledger Fabric guarantees data privacy, security, and interoperability, fostering seamless sharing of medical records among different healthcare providers. However, challenges include the complexity of implementing the network and the possible resistance from customary healthcare systems accustomed to centralized control. Careful consideration of these pros and cons is essential for successful adoption and performance in the healthcare industry. 
\subsection{Advantages of utilizing Hyperledger fabrics for Telemedicine Projects}
The advantages of using Hyperledger Fabric for healthcare projects are innumerable. It provides improved data privacy, security, and interoperability, enabling secure sharing of medical records. Additionally, Hyperledger Fabric's decentralized nature reduces dependency on arbitrators, encouraging efficiency and trust in the healthcare ecosystem. In the following, we present a set of pros as,
\begin{itemize}
    \item Enhanced Security and Privacy: Hyperledger Fabric offers a protected and private setting for telemedicine projects. It includes cryptographic algorithms and access controls to guarantee the confidentiality and integrity of susceptible patient data. The permissioned network permits only qualified participants to access and validate transactions, optimizing the risk of unauthorized access or data infringements.
     \item Scalability and Performance: Hyperledger Fabric is developed for scalability, authorizing telemedicine projects to take a large volume of transactions efficiently. It employs a modular architecture that isolates consensus and execution, enabling parallel processing of transactions. This characteristic guarantees high performance and responsiveness, even in scenarios with a large number of participants and contemporary interactions.

 \item Customizable Governance Model: Telemedicine projects often involve numerous stakeholders, such as healthcare providers, patients, insurers, and regulators. Hyperledger Fabric presents a adaptable governance model, letting organizations to define their own rules and guidelines. This customization allows the creation of tailored governance frameworks that align with the explicit prerequisites of telemedicine projects, cultivating trust and cooperation among participants.

 \item Immutable Audit Trail: Hyperledger Fabric sustains an unchangeable record of all transactions, forming a transparent and auditable track of events in telemedicine projects. This attribute is crucial for accountability, compliance, and dispute resolution. The capability to trace and confirm every transaction allows trust building among stakeholders, assuring the integrity of medical records, treatment history, and billing details.

 \item Interoperability and Integration: Telemedicine projects often involve incorporating numerous systems, such as electronic health record (EHR) systems, payment gateways, and IoT devices. Hyperledger Fabric supports interoperability by offering standardized APIs and protocols, enabling seamless integration with existing healthcare infrastructure. This ability enables telemedicine platforms to securely exchange data and interact with external systems, improving overall efficiency and usability.
\end{itemize}



\subsection{Disadvantages of utilizing Hyperledger fabrics for Telemedicine Project}
While Hyperledger Fabric offers numerous benefits for healthcare projects, there are certain disadvantages and challenges to consider.  In the following, we outline many such concerns and considerations that are crucial for blockchain technology's successful implementation in the healthcare industry.
\begin{itemize}
    \item Complexity and Learning Curve: Hyperledger Fabric is a sophisticated and complicated technology that demands specialized knowledge and expertise to execute and maintain. Designing and deploying a telemedicine project using Hyperledger Fabric may involve substantial learning curves for developers and managers who are unfamiliar with blockchain technology. This can raise the initial time and cost investments needed to establish the system and may pose challenges for institutions with little technical resources.

 \item Limited Transaction Throughput: While Hyperledger Fabric is designed for scalability, its consensus mechanism and modular architecture can even set limitations on transaction throughput. In telemedicine projects where real-time interactions and large volumes of data swap are paramount, the transaction processing capability of Hyperledger Fabric may fall short. Depending on the specific requirements, institutions may require to carefully assess the scalability and performance trade-offs to guarantee the system can endure the expected workload.

 \item Governance and Compliance Challenges: Hyperledger Fabric's customizable governance model, while advantageous in many regards, can also present governance and adherence challenges. In telemedicine projects, compliance with laws such as HIPAA (Health Insurance Portability and Accountability Act) is essential to safeguard patient privacy and confidentiality. Customizing governance guidelines and guaranteeing alignment with regulatory prerequisites can be complex and time-consuming. Institutions must carefully navigate these challenges to assure they stick to applicable laws and regulations while leveraging the advantages of Hyperledger Fabric.
\end{itemize}




\section{Conclusion}\label{conc}

The proposed blockchain-based framework has the potential to unlock great potential in healthcare by providing secure and accessible telemedicine and telehealth services. The Covid19 pandemic has highlighted the urgent need for such services, and blockchain technology can provide a secure and decentralized platform to facilitate remote medical consultations.
The inherent value of the Telehealth chain protocol will come from the fact that this blockchain protocol will allow patients in remote locations to handle certain functions, including recording their medical data into the blockchain, which in turn will be immediately available to the physicians, insurance company and hospitals, depending on who needs to respond to patient request. The protocols will be designed to also act as a payment service (using Hyperledger) and can also be used to pay for various services on the platform.

\bibliographystyle{unsrt}
\bibliography{mybibfile}
\end{document}